\documentclass[sn-basic,a4paper]{sn-jnl}

\usepackage{graphicx}%
\usepackage{multirow}%
\usepackage{amsmath,amssymb,amsfonts}%
\usepackage{amsthm}%
\usepackage{mathrsfs}%
\usepackage[title]{appendix}%
\usepackage{xcolor}%
\usepackage{textcomp}%
\usepackage{manyfoot}%
\usepackage{booktabs}%
\usepackage{algorithm}%
\usepackage{algorithmicx}%
\usepackage{algpseudocode}%
\usepackage{listings}%

\usepackage{ref_macros_sci}

\theoremstyle{thmstyleone}%
\theoremstyle{thmstyletwo}%

\theoremstyle{thmstylethree}%

\begin{document}

\title[Rise and shine]{Superluminous supernovae: diverse rise times explain diverse spectra}


%


\author*[1]{\fnm{Matt} \sur{Nicholl}}\email{matt.nicholl@qub.ac.uk}



\affil[1]{\orgdiv{Astrophysics Research Centre, School of Mathematics and Physics}, \orgname{Queen's University Belfast}, \orgaddress{\city{Belfast}, \postcode{BT7 1NN}, \country{UK}}}




\abstract{
Type I superluminous supernovae (SLSNe) are a diverse class of exceptionally bright massive star explosions, which typically exhibit absorption from ionised oxygen in their early spectra. While their photometric properties (luminosity and duration) both span an order of magnitude, population studies suggest that these distributions are continuous. However, spectroscopic samples have shown some indications of distinct sub-types, either through similarity to certain prototype objects, or in terms of their velocity evolution. Here we show that a well-observed SLSN, PTF12dam, completely changes its O II absorption profile as it rises to maximum light, moving from one proposed sub-type to another. This supports an interpretation where spectroscopic diversity is driven by the ejecta temperature at maximum light, rather than fundamental differences in the explosion or progenitor. Motivated by this, we develop a new diagnostic, the Brightness-Timescale-Temperature-Radius diagram, and a simple toy model for the evolution of the photospheric velocity, to show that diversity in the light curve rise time (likely due to differences in ejected mass) naturally explains why SLSNe with broader light curves generally have weaker O II lines, lower photospheric velocities after maximum, and slower changes in photospheric velocity over time. We show that the velocity distribution of the known SLSN population favours a relatively flat ejecta density profile, consistent with a hot bubble inflated by a central engine.
}




\maketitle

\section{Introduction}\label{sec:intro}

Hydrogen-poor superluminous supernovae (SLSNe I, or simply SLSNe) regularly reach peak luminosities $M<-21$\,mag, around $100\times$ brighter than a typical core-collapse supernova (SN). However, their modern definition does not use a hard luminosity cut. Instead the requirement is spectroscopic: a SLSN around maximum light exhibits a blue optical spectrum with absorption lines from ionised oxygen \citep{Quimby2011}. This spectroscopic net also catches some supernovae that are only marginally over-luminous \citep{DeCia2018,Lunnan2018,Angus2019,Gomez2022,Chen2022}. Yet regardless of the luminosity, the observed O II transitions require sustained high temperatures and possibly non-thermal excitation for several weeks during the rise to maximum light \citep{Mazzali2016,Dessart2019,Saito2024}. Thus the spectrum alone indicates an additional energy source compared to normal nickel-powered supernovae, making SLSNe of enormous physical interest \citep[e.g.][]{Gal-Yam2019,Inserra2019,Nicholl2021,Moriya2024}.

As a group they are diverse, not just in luminosity but in their durations and spectroscopic properties. Most appear to come from fully stripped stars similar to Type Ic SNe, but a fraction exhibit helium lines \citep{Yan2020,Kumar2025}. Evidence for pre-explosion mass-loss and circumstellar interaction also appears in the spectra of some SLSNe \citep[e.g.][]{Yan2017,Lunnan2018,Pursiainen2022,Aamer2024,Gkini2025}. The presence of interaction likely contributes to their diversity and possibly to their luminosity budget, and may help to explain the `bumps' and `wiggles' often detected in their light curves \citep{Leloudas2012,Nicholl2015a,Hosseinzadeh2022,Chen2022a}; however, see recent work by \citet{Farah2025} for an alternative explanation for these bumps.

A central energy source such as the spin-down of a nascent millisecond magnetar \citep{Kasen2010} can explain most of the photometric \citep{Inserra2013,Nicholl2017,Blanchard2020,Gomez2022} and spectroscopic \citep{Dessart2012,Mazzali2016,Jerkstrand2017,Aamer2025} properties of SLSNe, as well as naturally accounting both for their overall energetics and for their empirical connection to long-duration gamma-ray bursts \citep{Lunnan2014,Greiner2015,Nicholl2016}. However, with multiple energy sources potentially at play, it is feasible that Nature has several pathways to produce these explosions. Thus, a critical question is whether the SLSNe detected to date constitute one (rather diverse) population with a common origin, or several distinct groups with different progenitors and/or power sources.


It was first suggested that SLSNe may separate photometrically into events with fast and slow light curve evolution \citep{Gal-Yam2012,Inserra2018a}. This has been debated in the literature, with larger photometric samples showing no clear separation in timescales \citep{Nicholl2017,DeCia2018}. It has also been proposed that the SLSNe with slower light curves also exhibit lower photospheric velocities  and shallower velocity gradients at maximum light. An unsupervised clustering analysis in this space by \citet{Inserra2018a} favoured two sub-populations, though the sample available at that time comprised only 12 events.

The profiles and equivalent widths of certain spectral lines have also been observed to vary between SLSNe. \citet{Quimby2018} suggested two distinct groupings based on template matching of photospheric-phase spectra. However, \citet{Nicholl2019} argued that there were no statistical differences between their spectra once these events reached the nebular phase. Another study by \citet{KonyvesToth2022} suggested dividing SLSNe into two groups based on the profiles of their O II lines, but again the differences between these groups seemed to disappear by the late photospheric and early nebular phases \citep{KonyvesToth2023}. 

The extensive SLSN Catalog recently compiled by \citet{Gomez2024}, containing over 200 events, does not exhibit strong evidence for SLSN sub-classes; rather it appears to show a continuum of light curve timescales \citep{Gomez2024} and spectroscopic properties \citep{Aamer2025}. But these large samples do still reveal a very diverse population. Therefore even if SLSNe constitute a continuous group, the most interesting questions still remain: what are the \textit{physical} factors that lead to the observed diversity of SLSNe, and do the same factors account for both their photometric and spectroscopic diversity?

In this paper, we argue that the spectroscopic diversity of SLSNe at maximum light can be explained in large part by their wide range of light curve rise times, which span a much wider range than other SN classes. We first revisit the pre-maximum spectra of the SLSN PTF12dam \citep{Nicholl2013,Vreeswijk2017,Quimby2018}, and show that the evolution of its O II line profile defies proposed classification schemes and suggests that SLSNe broadly follow a common spectral evolution after explosion. We then verify the importance of the rise time in determining the spectrum \emph{observed at maximum light}, using a new diagnostic tool for transients: an extension of the common luminosity-duration phase space that now also includes the ejecta temperature and radius. Finally, we show using a simple toy model that the observed velocities and velocity gradients in SLSNe can also be explained by a single population with a common density profile (but a wide range of ejecta masses) sampled at different times relative to explosion. Together, these considerations indicate that one parameter -- the light curve timescale -- can naturally account for much of the variation seen in SLSN spectra at and after peak.

\begin{table}[b]
\centering
\begin{tabular}{cccc}
SLSN & \citet{KonyvesToth2022} &  \citet{Quimby2018} & \citet{Inserra2018a}  \\
\hline
PTF09cnd & W & 12dam-like & Fast \\
SN2015bn & 15bn-like & -- & Slow \\
SN2011ke & -- & 11ke-like & Fast \\
PTF12dam & W & 12dam-like & Slow \\
\hline
\caption{Summary of sub-class assignment in the literature for the key SLSNe that define the different classification schemes. Note that not all events have been included within every scheme, due to data availability / sample selection in these studies.}
\end{tabular} 
\label{tab:classes}
\end{table}

\section{O II line profiles and PTF12dam}\label{sec:oii}

\subsection{Proposed spectroscopic sub-types}

We first expand on the different SLSN spectroscopic sub-types, based on similarity to prototype events, proposed in the literature. For ease of reference, we provide a matrix of these grouping schemes and the prototypical events for each group in Table \ref{tab:classes}. We also indicate whether each event belongs to the proposed `fast' and `slow' photometric sub-types.

\textbf{`W' vs `15bn':} \citet{KonyvesToth2022} and \citet{KonyvesToth2023} defined two possible sub-classes of SLSNe based on the shapes and ratios of their O II absorption lines -- in particular the two reddest and strongest blends with rest-frame wavelengths of 4651\,\AA\ and 4358\,\AA\ \citep{Quimby2018}. With typical velocities $>10,000$\,km\,s$^{-1}$, these lines are usually blueshifted to $\approx 4100-4600$\,\AA. In the proposed `W' class, the two lines have approximately equal strengths, and each has a roughly triangular profile. In their `15bn-like' class, the reddest line is weaker, and the lines have more complex profiles with flatter troughs or multiple minima. A typical example of each group is shown in Figure \ref{fig:evol}. For the 15bn-like group, we show the prototype, SN2015bn at phases within $\pm10$ days of maximum light \citep{Nicholl2016a,Nicholl2016,Nicholl2018}. To represent the `W' profile, we show the spectrum of PTF09cnd at 21 rest-frame days before maximum light \citep{Quimby2011}. SN 2015bn is classed by \citet{Inserra2018a} as a slow event, and PTF09cnd as a fast event.

\textbf{`PTF12dam' vs `SN2011ke':} \citet{Quimby2018} studied a sample of SLSNe from the Palomar Transient Factory, and defined a spectroscopic `phase' based on the typical sequence evolving from the hot O II phase, through the cooler photospheric phase, to the nebular phase. They used the same spectrum of PTF09cnd shown in Figure \ref{fig:evol} to define the earliest spectroscopic phase, when O II is strongest. Comparing spectra in similar phase bins, they identified two possible groups, with one defined by similarity to SN2011ke \citep{Inserra2013}, and the other defined by similarity to PTF12dam \citep{Nicholl2013,Vreeswijk2017}. This similarity was established by comparing the ranks of best matching templates with the code \textsc{superfit} \citep{Howell2006}. We include a spectrum of SN2011ke at 8 days after maximum in Figure \ref{fig:evol} (and discuss PTF12dam in detail below). \citet{Inserra2018a} classify PTF12dam as a slow event, and SN2011ke as a fast event.

\subsection{The spectroscopic evolution of PTF12dam}

PTF12dam is a SLSN with a $\sim 50$\,day rise and slow decline. It has an excellent time-series of spectra beginning before maximum light, exhibiting strong O II lines at early phases, later replaced by Fe II and Fe III \citep{Nicholl2013}. As discussed, it defines one of the groups proposed by \citet{Quimby2018}. It was also studied by \citet{KonyvesToth2022}, who included it in their `W' class. PTF12dam represents an interesting case study, since it has been classed as a slowly-evolving event but is spectroscopically distinct from the even slower SN2015bn at maximum light\footnote{However, note \citet{KonyvesToth2023} found that the \emph{post}-maximum spectra of PTF12dam and SN2015bn could no longer be cleanly separated.}. It is also distinct from SN2011ke, which lacks a pre-maximum spectrum but would be expected to belong to the `W' group, since all events deemed to be `fast' fall in that class. Therefore, taken together these events immediately demonstrate that there is no simple mapping between the \citet{Quimby2018}, \citet{KonyvesToth2022} and \citet{Inserra2018} sub-classification schemes, as shown in Table \ref{tab:classes}.

\begin{figure*}[t!]
\includegraphics[width=\textwidth]{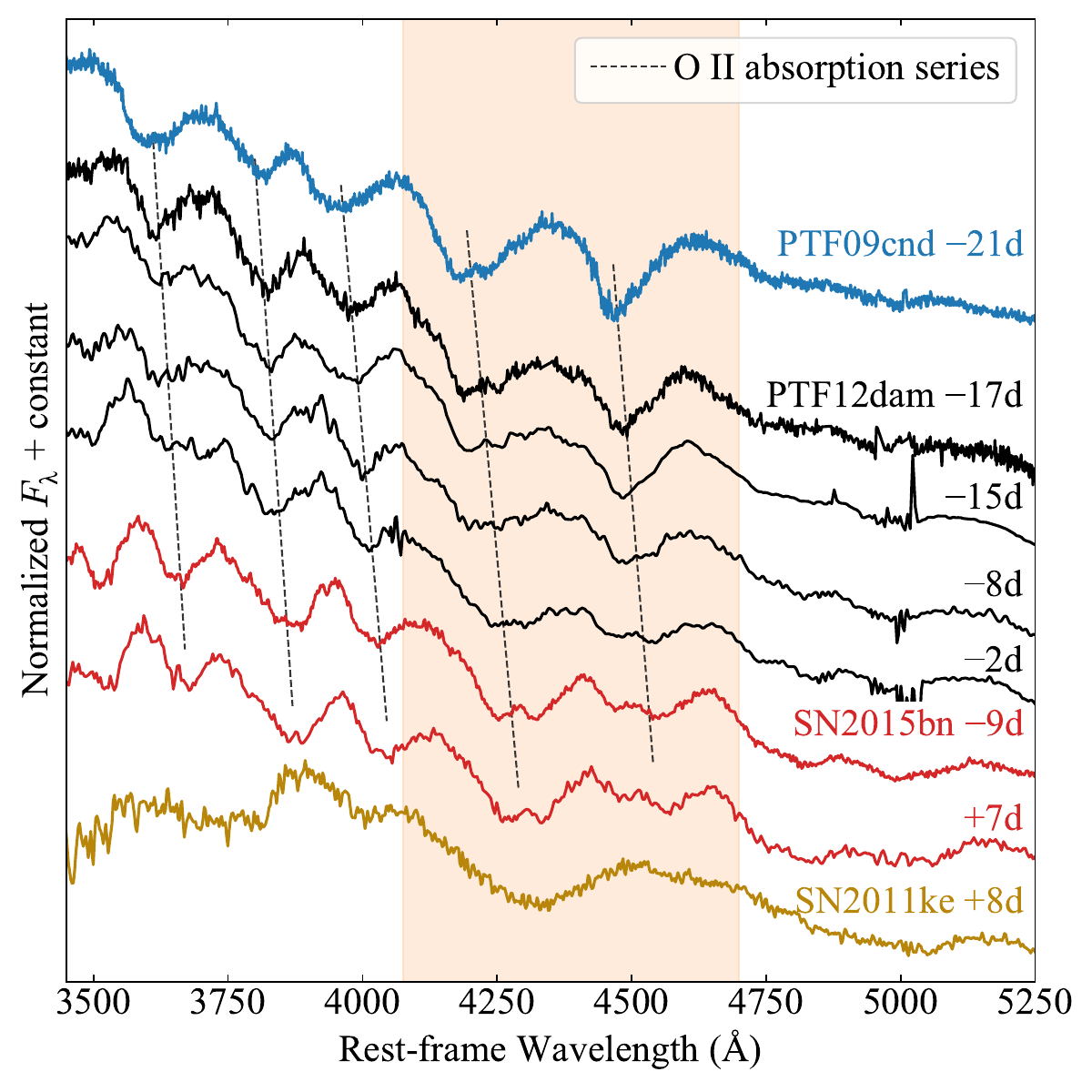}
\caption{O II lines in the early spectra of SLSNe defining the spectroscopic classes proposed in the literature (Table \ref{tab:classes}). Phases are given in rest-frame days from maximum light. The shaded region highlights the strongest two lines. The O II line lines in PTF12dam \citep{Nicholl2013,Vreeswijk2017} transition smoothly between the `W'-like profiles exemplified by PTF09cnd \citep{Quimby2011} and the more complex profiles exhibited by SN2015bn \citep{Nicholl2016a}. SN2011ke \citep{Inserra2013}, which has a much shorter rise time than the others, exhibits no O II lines by $\sim1$ week after maximum, as well as much higher velocities.
\label{fig:evol}}
\end{figure*}

We retrieve from \citet{Nicholl2013} and \citet{Vreeswijk2017} the four pre-maximum spectra of PTF12dam with the greatest signal-to-noise ratios. This event occurred in an extreme starbursting galaxy at $z=0.107$, and thus the spectra are heavily contaminated by galaxy lines. We remove these using Gaussian fits, as in \citet{Aamer2025}. We plot the resulting spectra in Figure \ref{fig:evol} alongside SN2015bn, PTF09cnd and SN2011ke. Each spectrum has been normalised by its mean value, and offset vertically for clarity, and the (blueshifted) positions of the five O II absorption lines are marked. 

Plotted in this way, it can be readily observed that the earliest spectra of PTF12dam closely match PTF09cnd, with a pronounced `W' shape between 4100-4650\,\AA. However, over the following weeks, both the velocities and shapes of these lines undergo a substantial change. By the time of maximum light, the spectra of PTF12dam provide a much better match to the pre-peak spectra of SN2015bn; visual inspection would almost certainly place it in the 15bn-like class. As noted by \citet{Quimby2018}, the earliest spectrum of SN2011ke (though notably, obtained more than a week past maximum light) displays a somewhat different character. While the overall shape is more similar to SN2015bn, the lines are much broader (i.e. form at higher velocities) and are likely dominated by Fe II at this phase \citep{Inserra2013}. We suggest a reason for these differences in section \ref{sec:bttr}.

\begin{figure*}[t!]
\includegraphics[width=\textwidth]{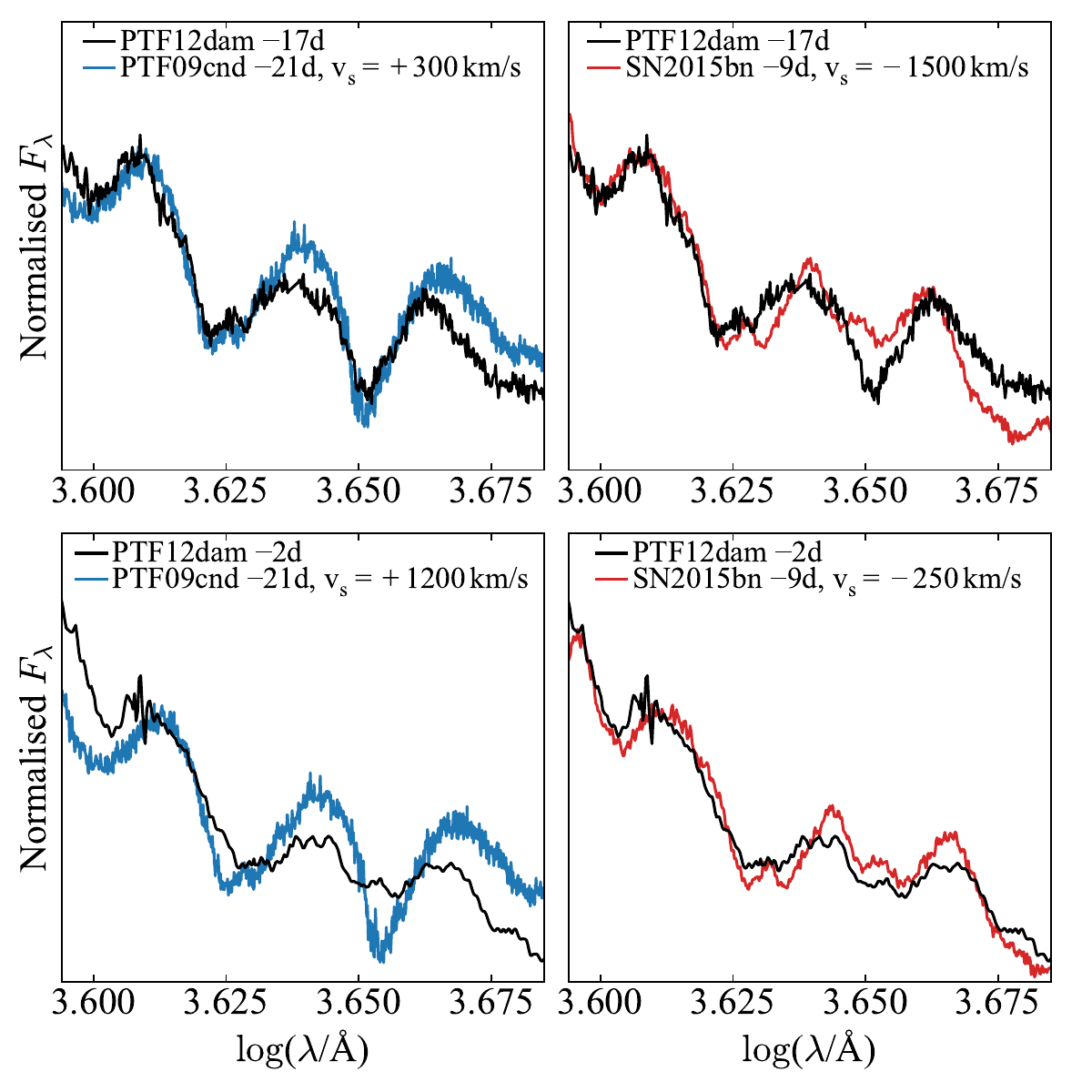}
\caption{Zoom-in around the two strongest O II lines in PTF12dam, compared to PTF09cnd and SN2015bn. The comparison spectra are shifted along the logarithmic wavelength axis to account for velocity differences \citep{Quimby2018}. This shows clearly that while the earliest spectra of PTF12dam have a strong `W' feature, the later spectra are much closer in character to SN2015bn.
\label{fig:zoom}}
\end{figure*}

Returning to PTF12dam, we show in Figure \ref{fig:zoom} a more direct comparison of the O II line profiles with PTF09cnd and SN2015bn. Following \citet{Quimby2018}, we plot these spectra on a logarithmic wavelength scale, such that velocity differences can be accounted for by simple translations along this axis. In the top row, we show the spectrum of PTF12dam at $-17$\,days. Redshifting the spectrum of PTF09cnd by a modest 300\,km\,s$^{-1}$ gives an excellent match to the `W' feature, including the slight `double-dip' in the bluer of the two lines. SN2015bn has much lower line velocities, but blueshifting its spectrum by 1500\,km\,s$^{-1}$ gives a reasonable match to the positions of the O II lines. However, the profiles and ratios are clearly discrepant. In the bottom row, we show the spectrum of PTF12dam at $-2$\,days. In this case, the deep absorptions in PTF09cnd provide a poor match to the shallower, asymmetric absorptions now seen in PTF12dam, even after redshifting PTF09cnd by 1200\,km\,s$^{-1}$ to match the velocities. SN2015bn provides a much closer match to the line ratios and profiles at this phase, and in particular to the complex profile of the reddest O II line, with only a small velocity shift.

From this analysis it would appear that SLSN spectra cannot be divided neatly into `W' or 15bn-like groups even prior to maximum light: the profile observed depends on the time of the observation. This is perhaps unsurprising: \citet{KonyvesToth2022} found that the differences in line profiles were closely linked to the photospheric temperatures in their sample, with only events hotter than $\approx12,000$\,K showing deep W-shaped profiles. This strong temperature dependence was also recovered theoretically in radiative transfer modelling by \citet{Saito2024}. If a given SLSN is observed closer to the time of explosion, it is likely to be hotter (under the reasonable assumption that an initially hot ejecta usually cools monotonically) and therefore more SLSNe are likely to show the `W' at earlier phases.

\section{Brightness-Timescale-Temperature-Radius (BTTR) analysis}\label{sec:bttr}

Since a SLSN evolves from a `W' to a 15bn-like profile as it cools, the character of its spectrum at the time of maximum light depends on its temperature during this snapshot in time. Since SLSNe exhibit some spread in temperature at maximum light, they will appear different at this phase, even if they might have looked similar at earlier times. We now posit that a wide range in photospheric temperature at maximum light is a natural (probably unavoidable) property of any transient population with intrinsic diversity in rise times and/or heating rates. SLSNe meet both of these criteria.

To see this clearly, we (re)introduce a useful diagnostic plot for understanding supernova diversity (an earlier version of this plot with 38 SLSNe was first presented by \citealt{Nicholl2017}). This visualisation tool was inspired by the classic `luminosity-duration' phase space, commonly used in the literature since \citet{Kulkarni2007}. We extend this parameter space to also include the photospheric temperature and radius at maximum light (or in principle at any desired phase). Thus a single plot intuitively shows the distributions of brightness, timescales, temperatures and radii, and the relationships between them, for any transient population. We therefore refer to this a `BTTR' diagram.

Obviously, these parameters are not completely independent, with any two of the luminosity, temperature and radius determining the third via the Stefan-Boltzmann law. However, plotting these quantities together is very advantageous for building physical insight. This method shows, for example, whether temperature differences between two objects are due to intrinsic differences in luminosity (roughly equivalent to the heating rate at maximum; \citealt{Arnett1982}) or because one is simply more expanded than the other. The time axis shows whether a larger radius occurs because more time has elapsed since explosion, or because of a higher velocity.

\begin{figure*}[t!]
\includegraphics[width=\textwidth]{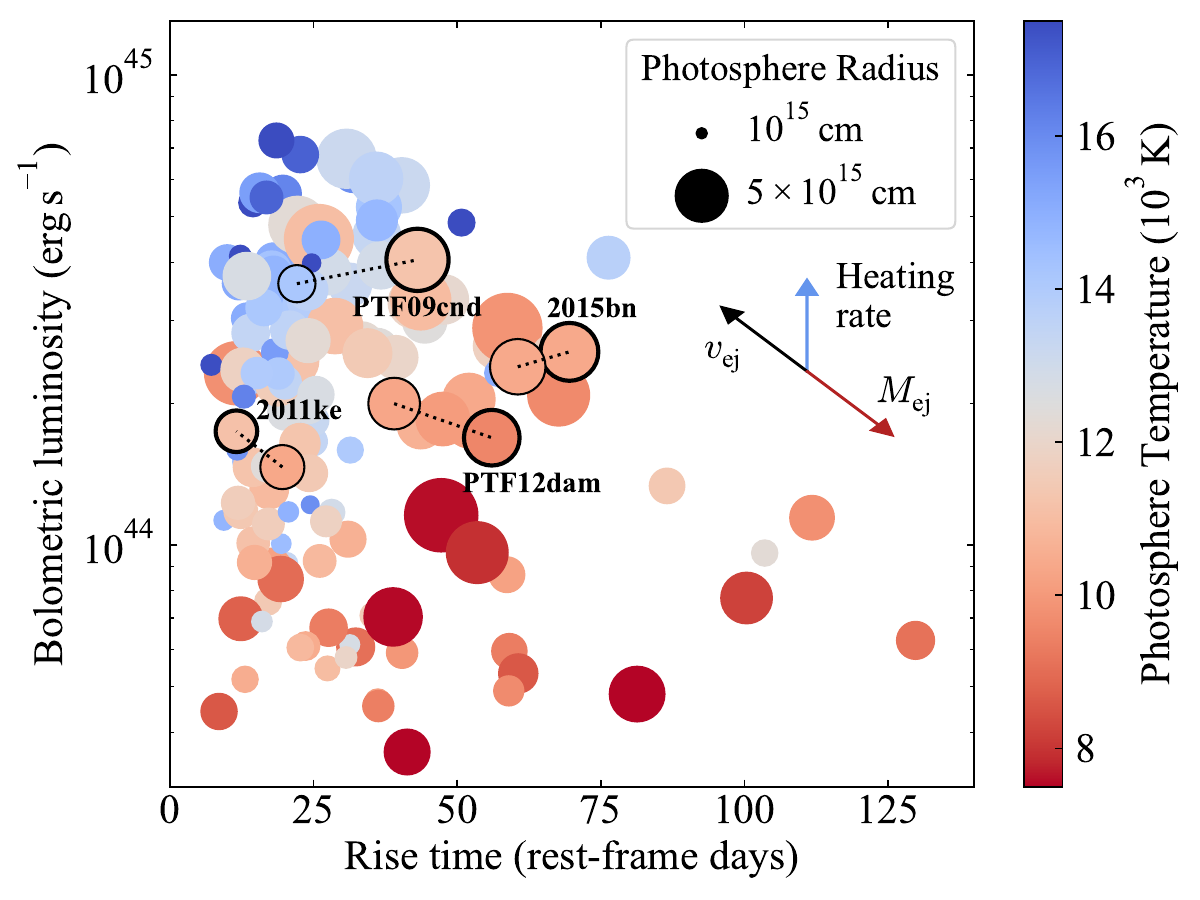}
\caption{Brightness-Timescale-Temperature-Radius (BTTR) diagram for all objects in the SLSN Catalog \citep{Gomez2024}. For most objects we plot a single point corresponding to the peak of its bolometric light curve. The SLSNe in Table \ref{tab:classes} are highlighted with a bold outline. For these four events, we plot an additional point with a narrow outline, corresponding to the time of the earliest spectrum plotted in Figure \ref{fig:evol}. Arrows show the effects of increasing the values of key underlying physical parameters: ejecta mass, velocity, and input heating rate. The colour of each arrow indicates the effect on the observed temperature.
\label{fig:bttr}}
\end{figure*}

We construct a BTTR plot for SLSNe using the SLSN Catalog \citep{Gomez2024}, which provides bolometric light curves and photospheric blackbody parameters for 143 SLSNe, derived using \textsc{extrabol} \citep{Thornton2024}. Figure \ref{fig:bttr} shows that SLSNe span roughly an order of magnitude in both their peak luminosities and rise times. However, the parameter space is not uniformly populated, and several trends are immediately clear. The bulk of SLSNe have rise times less than $\sim 50$\,days. No object with a rise time $>75$\,days has a luminosity above $\approx10^{44}$\,erg\,s$^{-1}$, though many of the faster objects have higher luminosities. This likely reflects the difficulty of sustaining a high degree of energy input over such long durations. Another clear trend is that events with longer rise times have larger maximum-light radii and are relatively cool. 

We highlight the effects of the underlying explosion parameters on the figure with schematic arrows. Increasing the heating rate from the power source, with all else equal, leads to brighter and bluer peaks; this follows from Arnett's rule (the luminosity at peak is roughly equal to the instantaneous heating rate). Larger ejecta mass ($M_{\rm ej}$) leads to a longer photon diffusion time and a later peak, by which time the instantaneous heating rate is lower and the ejecta are more expanded; this results in fainter, redder events. The effect of the photospheric velocity is more subtle: a larger velocity decreases the photon diffusion time, leading to earlier and brighter peaks, but the colour is determined by the competing effects of the higher luminosity (bluer for a given radius) and the larger photospheric radius (redder for a given luminosity) resulting from the rapid expansion.

We mark on this diagram the positions of the four SLSNe from section \ref{sec:oii}, showing them both at maximum light and at the times their earliest spectra were obtained. It is apparent that PTF12dam evolves over this interval from the transitional region between hotter and cooler events, and by the time of maximum light (when it resembles SN2015bn) it has firmly reached the parameter space occupied by the more expanded, cooler SLSNe. PTF09cnd, at the time of the first spectrum, is clearly in the high-temperature regime, whereas SN2011ke is already rather cool by the time a spectrum was obtained. The BTTR plot indicates that this is because SN2011ke has a modest luminosity (by the standards of this class), rather than an especially extended photosphere. 

This analysis suggest that SLSNe have a wide (but continuous) distribution in rise times and heating rates, and it is the interplay between these characteristics (rather than a fundamentally different ejecta or explosion mechanism) that determines when and how the O II lines appear. However, one aspect we have not considered in our analysis so far is the observed correlation in spectral line velocities and velocity gradients \citep{Inserra2018a}. We next ask whether the range of SLSN rise times can explain this phenomenon also.

\section{Constraining the density structure of SLSNe}\label{sec:velocities}

The early spectrum of any supernova (or expanding thermal transient) begins with blackbody emission from the photosphere (the `surface' inside the ejecta where the optical depth to an observer is $\tau\sim1$). As this radiation passes through the outer atmosphere, atomic transitions are imprinted at a range of velocities. Typical supernova ejecta expand homologously ($v\propto r$, where $r$ is the radial coordinate inside the ejecta). Since we can only observe spectral lines from material outside of the photosphere, the observed Doppler velocities of spectral lines follow $v_{\rm obs}\geq v_{\rm ph}$, the instantaneous photospheric velocity. As the ejecta expand and more of it becomes optically thin, the photosphere recedes deeper in the homologous flow, such that lower line velocities are observed as time increases. Absorption from the Fe II $\lambda5169$ multiplet is thought to saturate close to the photosphere, and hence it is often used as a tracer of $v_{\rm ph}$ \citep[e.g.][]{Modjaz2016}. 

It was noted by \citet{Inserra2018a} that SLSNe with broader light curves tended to exhibit lower $v_{\rm ph}$, as measured by Fe II, after maximum light. Such events also showed more gradual changes in apparent line velocity ($\dot{v}\equiv{\rm d}v_{\rm ph}/{\rm d}t$) at the same phase, leading to a strong correlation between $v_{\rm ph}$ and $\dot{v}$. Using a larger sample from the SLSN Catalog, \citet{Aamer2025} reproduced this correlation, but did not find a clear gap between `fast' and `slow events' (in this case, fast and slow refers both to the observed velocities and to the light curve timescales).

We showed in the previous section that long-rising (slow) events are naturally expected to be cooler at maximum light (for a given luminosity). Since velocity evolution is also commonly anchored relative to maximum light, it is logical to ask whether there is another physical effect that could explain why events observed at quite different times \emph{relative to explosion} might show systematic differences in $v_{\rm ph}$ and $\dot{v}$. To investigate this, we apply a simple analytic model to SLSN velocity data to test whether a single density profile, observed at different phases from explosion, can reproduce the observations.

\subsection{Analytic model for the photospheric velocity}

Our model is defined as follows. The ejecta expand homologously, with a density profile that depends on velocity coordinate as:
\begin{eqnarray}\label{eq:density}
\rho(v) = 
\left\{
\begin{array}{lr}
 \rho_0 ,\ &
 v\leq v_{\rm br} \\
 \rho_0(v/v_{\rm br})^{-\alpha},&
 v> v_{\rm br}\\
\end{array}
\right.
\end{eqnarray}

where $v_{\rm br}$ is a characteristic break velocity between a flat inner part (a dense core) and a steep outer envelope. For a given $M_{\rm ej}$ and $v_{\rm br}$, the normalisation $\rho_0$ is determined by the condition

\begin{equation}
    M_{\rm ej} = \int_0^{\infty} \rho {\rm d}V = \int_0^{\infty} \rho(v) 4 \pi v^2t^3{\rm d}v,
\end{equation}

where the latter equality follows from the volume below velocity coordinate $v$, given by $V=4 \pi (vt)^3 /3$. At a given time $t$, the velocity of the photosphere can be determined by integrating the optical depth seen by an observer, ${\rm d}\tau= \kappa \rho(r,t) {\rm d}r$, and using ${\rm d}r = t{\rm d}v$ to solve
\begin{equation}
    \int_\infty^{v_{\rm ph}} \kappa \rho(v,t) t {\rm d}v =1.
\end{equation}
This gives an analytic expression for the velocity, with time evolution that follows
\begin{equation}
    v_{\rm ph} \propto t^{-2/(\alpha - 1)},
    \label{eq:v}
\end{equation}
for $v_{\rm ph}>v_{\rm br}$. Differentiating this with respect to time yields
\begin{equation}
    \dot{v} \propto t^{-(\alpha+1)/(\alpha - 1)}.
    \label{eq:dv}    
\end{equation}


\begin{figure*}[t!]
\includegraphics[width=0.5\columnwidth]{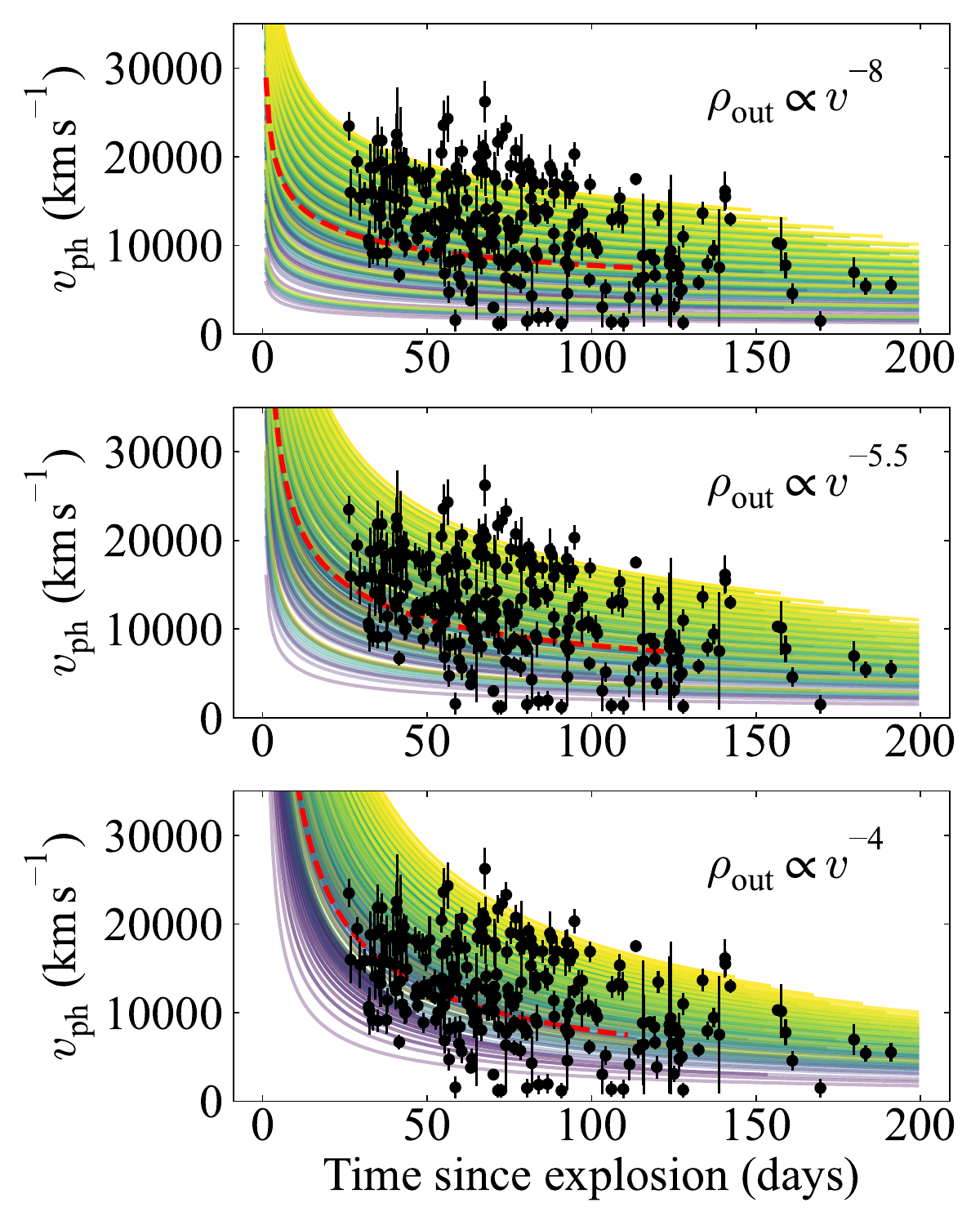}
\includegraphics[width=0.5\columnwidth]{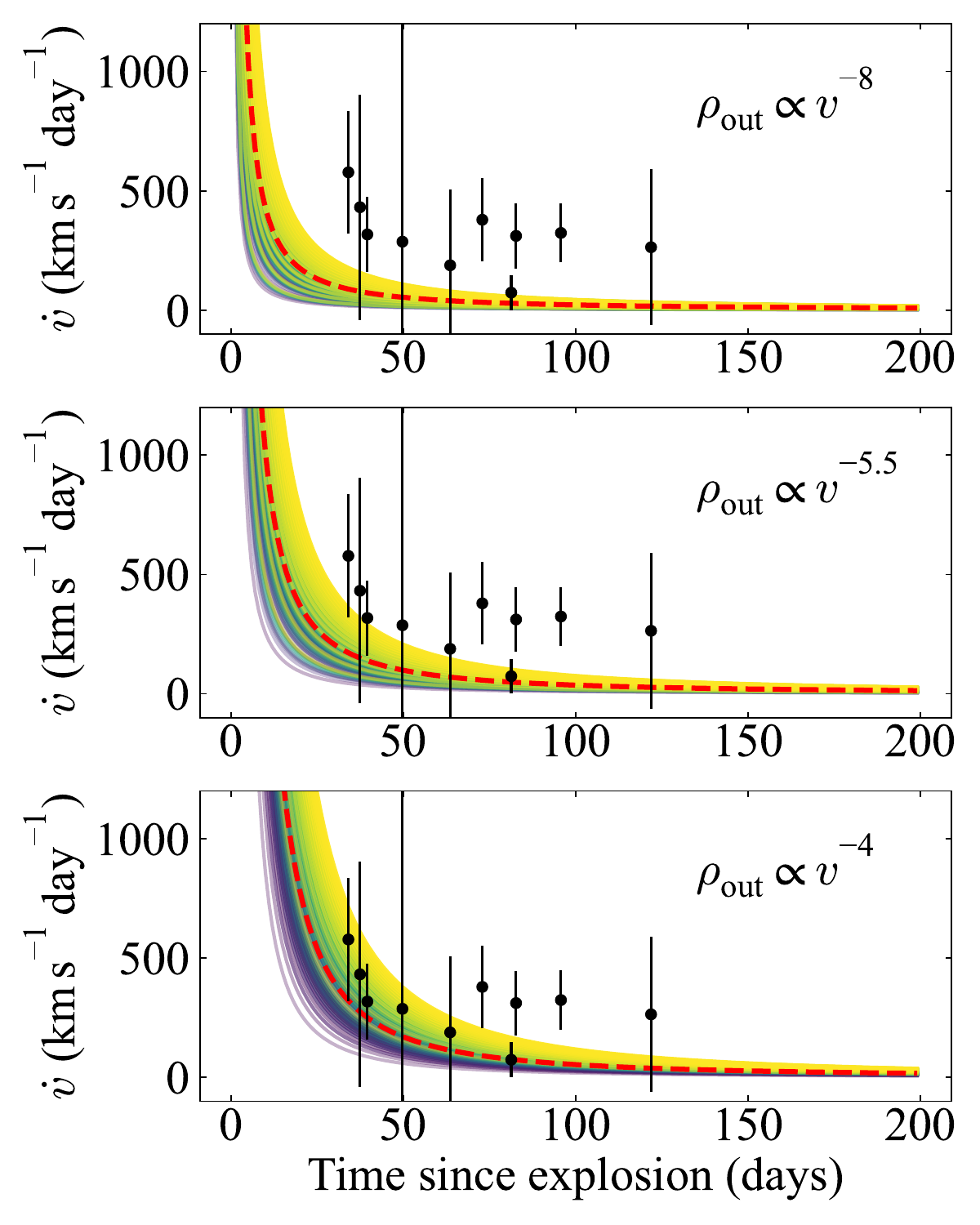}
\caption{Analytic models for the time-dependence of $v_{\rm ph}$, compared to the observed population of SLSNe from \citet{Aamer2025}. The left column shows $v_{\rm ph}$, and the right shows its time derivative, $\dot{v}$. Each row corresponds to a different slope of the ejecta density profile, as labelled. The colour scale corresponds to different ejecta masses, ranging from 2\,M$_\odot$ (purple) to 40\,M$_\odot$ (yellow) to cover the full range of masses reported by \citet{Gomez2024}. The red dashed line uses their median inferred ejecta mass and kinetic energy. 
\label{fig:v}}
\end{figure*}

\subsection{Application to SLSNe}

We retrieve the velocities $v_{\rm ph}$ and velocity gradients $\dot{v}$ from the SLSN Catalog \citep{Aamer2025}, and plot these in Figure \ref{fig:v}. We compare these to the predictions from a grid of simulations using the analytic model described above. Each model is terminated once $v_{\rm ph}=v_{\rm br}$ and equations \ref{eq:v} and \ref{eq:dv} no longer apply; in practice, this condition is reached by only a few relevant models, and only on timescales of several months after explosion.

Specifically, we vary the ejecta mass over the full range of masses from $2\leq M_{\rm ej}\leq40\,{\rm M}_\odot$, derived from light curve modelling by \citet{Gomez2024}, and adopt a wide range of plausible break velocities between $1000\leq v_{\rm br}\leq 15000$\,km\,s$^{-1}$. We also compute a model for the `median' SLSN from the SLSN Catalog, with $M_{\rm ej}=6.5$\,M$_\odot$ and kinetic energy $E_{\rm k}=2.5\times10^{51}$\,erg \citep{Gomez2024}. The break velocity can be estimated using $v_{\rm br} \approx 3500 \sqrt{E_{51}/M_{10}}$\,km\,s$^{-1}$, where $E_{51}$ and $M_{10}$ are the kinetic energy and ejecta mass in units of $10^{51}$\,erg and 10\,M$_\odot$, respectively \citep{Suzuki2019}. This gives a break velocity of $\sim7400$\,km\,s$^{-1}$ for the median SLSN (this also motivates the range of $v_{\rm br}$ used for our model grid).

We show results for three physically-motivated choices of the density slope parameter, $\alpha$. We first show a steep profile with $\alpha=8$, motivated by classic models of stripped-envelope supernovae \citep[e.g.][]{Iwamoto2000}. This model underpredicts the velocities of SLSNe, as the photosphere recedes too quickly through the low-density outer envelope and quickly reaches the slower inner regions. The velocity gradient is quite flat after $\sim 1$ month from explosion, as the photosphere now recedes slowly in the dense inner ejecta, and is not consistent with the observations.

We have more success in matching observations when using a flatter density profile. We next choose the value $\alpha=5.5$ based on the three-dimensional SLSN simulations from \citet{Suzuki2019}. They found that SLSN ejecta heated and accelerated by a central magnetar engine naturally reached a phase of homologous expansion, with a density profile exponent between $5\lesssim\alpha\lesssim6$. This shallower profile is the result of the pressure exerted on the ejecta by a hot inner bubble, inflated by the magnetar. This bubble also formed in two-dimensional simulations by \citet{Chen2016} and \citet{Suzuki2021}. Using this exponent in our model, we broadly recover the observed range of velocities, though we also predict that many SLSNe should have velocities $<10,000$\,km\,s$^{-1}$ at $\sim 1$ month after explosion, in slight tension with observations. This model also appears to be marginally consistent with the observed $\dot{v}$, though the predictions are systematically on the low end compared to observations.

For the final model, we use a very shallow slope, $\alpha=4$, corresponding to the model used by \citet{Mazzali2000} to fit the spectrum of the energetic broad-lined SN Ic, 1997ef. This model provides excellent agreement to the $v_{\rm ph}$ distribution of the SLSN population at all phases from explosion, and provides a good match to the early $\dot{v}$. This supports the need for a flatter density profile in SLSNe compared to expectations for typical supernovae \citep[e.g.][]{Matzner1999}, and is even shallower (but still conceptually similar) compared to the 3D magnetar simulations of \citet{Suzuki2019}. As noted by \citet{Maeda2023}, earlier observations will be important in more accurately measuring this density slope and constraining the ejecta and engine properties. This is apparent in Figure \ref{fig:v}, where the largest differences between the predictions for different models is within the first $\sim50$\,days from explosion.

\citet{Suzuki2019} noted that spectroscopic diversity within the SLSN population, including the existence of different sub-classes, could arise if a flat density profile leads to apparent high velocities in a sub-set of events. They proposed that in events with faster light curves (lower ejecta masses), this effect would be more pronounced, as the dynamics of lower-mass ejecta are more sensitive to the central engine; this could explain why events with fast light curves also tend to have higher photospheric velocities. Here, we have identified an additional effect that compounds this: it is not just the physical differences in ejected mass, but also the observational differences in when spectra are obtained relative to explosion (a point also emphasised by \citealt{Aamer2025}), that determine the observed $v_{\rm ph}$. When accounting for this latter effect, the observed distribution of Fe II velocities is consistent with a common, shallow ejecta density profile.

\begin{figure*}[t!]
\includegraphics[width=\columnwidth]{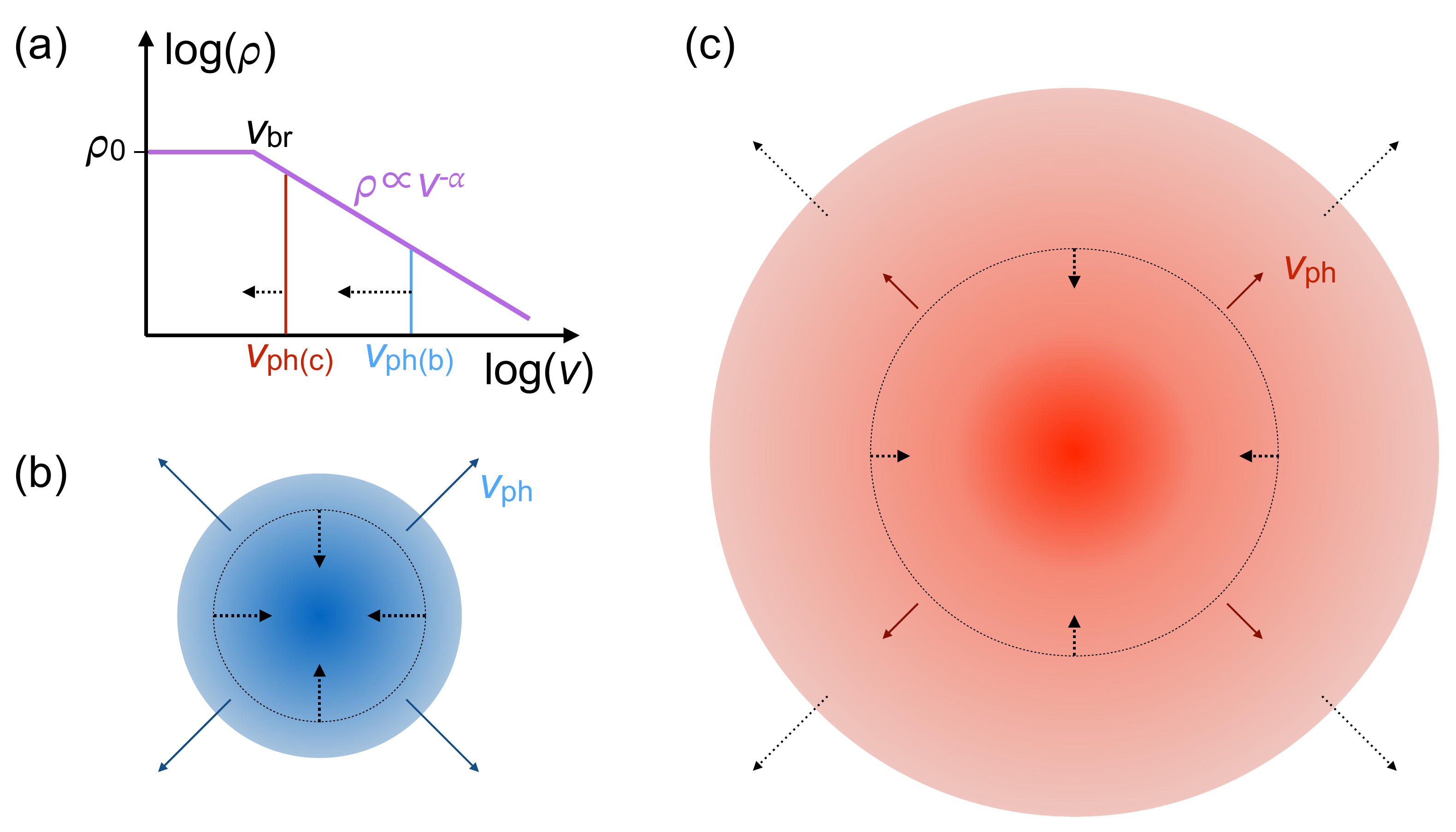}
\caption{Schematic showing how the observed spectroscopic properties of a SLSN at peak depend on its rise time. The relative lengths of all arrows are indicative of the relative velocities of expansion (solid arrows) and photospheric recession (bold, dashed arrows). \textbf{(a)} The density profile from equation \ref{eq:density}. We assume the ejecta are in homologous expansion, with a dense inner region and a shallow power-law envelope ($\rho\propto v^{-\alpha}$, with $\alpha<6$; see Figure \ref{fig:v}). The location of the photosphere is marked for the two cases shown in panels (b) and (c). \textbf{(b)} For a SLSN with a short rise time, the ejecta are still compact at maximum light, and therefore hotter for a given luminosity. For temperatures $\gtrsim12,000$\,K, deep `W'-shaped O II lines are produced. The photosphere (dotted line) is at high velocity coordinate, but receding quickly through the low-density outer ejecta; an observer therefore measures both a high $v_{\rm ph}$ and a large $\dot{v}$. \textbf{(c)} For a SLSN with a long rise time, the ejecta have expanded to a larger radius by maximum light, and so are cooler for the same luminosity. The O II lines may have already become weak or absent by this phase. The photosphere has receded further in mass coordinate by this time, and its rate of recession has slowed as it moves through increasingly dense ejecta. An observer therefore measures a lower $v_{\rm ph}$ and $\dot{v}$ compared to the fast-rising case.
\label{fig:schematic}}
\end{figure*}

We show in Figure \ref{fig:schematic} a simple schematic to emphasise the connection between this result and the findings in the previous sections. The extent to which the ejecta have expanded by maximum light determines both the temperature (for a given luminosity; section \ref{sec:bttr}) and the density. The former determines which spectral lines appear, the latter determines how far and how fast the photosphere can recede. This picture naturally explains why SLSNe with longer rise times (larger ejected masses) are more likely to show \emph{both} `15bn-like' (cooler) spectra, and lower velocities and velocity gradients.

\section{Conclusions}

We have conducted an analysis into the factors that affect SLSN spectroscopic diversity at maximum light, motivated by the observed evolution of the O II lines in PTF12dam from a `W'-shaped profile to a `15bn-like' one over the weeks leading up to maximum light. This is the first time such a clear transition has been demonstrated in an individual SLSN, and suggests that we should be cautious in ascribing sub-classifications to these events, since this classification is inherently time-dependent. In agreement with previous studies \citep{KonyvesToth2022,Saito2024}, we find that the ejecta temperature at the time of observation is most likely responsible for determining whether or not a strong `W' profile is apparent.

We then showed that the wide range of rise times in SLSNe naturally leads to a wide range of temperatures (and hence diversity in O II line profiles) at maximum light. We introduced a novel diagnostic, the BTTR diagram, to simultaneously probe the distributions of luminosities, rise times and photospheric blackbody parameters. This demonstrated that events with long rise times are cooler at maximum for a given luminosity because their photospheres have had more time to expand. The longest-rising events also tend to have a lower bolometric luminosity at peak, since sustaining a large heating rate over an extended period would challenge the energy budget even of most SLSNe. Defining spectroscopic phase with respect to maximum light (rather than time of explosion) is observationally convenient, but it obscures this physical connection between the rise time and spectrum formation.

We then asked whether differences in rise times could also explain why events with slower light curves often exhibit lower velocities and flatter velocity evolution in the weeks after maximum light. Using a toy model of a receding photosphere in a homologously expanding ejecta with a power-law density profile, we were able to reproduce the observed velocities from the SLSN Catalog \citep{Aamer2025}. The physical reason is that in events with longer rise times, the photosphere has had more time to recede in mass coordinate, and so has reached a region of the ejecta with a lower velocity and a higher density (leading to slower recession). Interestingly, we find that the data are best matched by a rather flat density profile, consistent with predictions from multi-dimensional simulations with central engine powering \citep{Suzuki2019}.

Overall this study shows that the spectroscopic diversity of SLSNe, in terms of both their O II lines and velocity evolution, is intimately connected to the diversity in their rise times, and that this connection results from basic physical considerations, without requiring distinct sub-classes. As discussed extensively in the literature, differences in rise times can be driven mainly by differences in ejected mass. Since several population studies have shown a continuous distribution of ejecta masses in SLSNe \citep{Nicholl2017,Blanchard2020,Chen2022a,Gomez2024}, this produces a continuum of O II line strengths and photospheric velocities, rather than distinct sub-populations. We expect that the methods developed here will be useful diagnostics for analysing the thousands of SLSNe expected from the imminent Rubin Observatory Legacy Survey of Space and Time \citep{Villar2018}.


\backmatter

\bmhead{Acknowledgements}

MN is grateful to the Fondation MERAC and to the organisers/participants of Superluminous 2025.

\section*{Declarations}


\begin{itemize}
\item Funding: MN is supported by the European Research Council (ERC) under the European Union’s Horizon 2020 research and innovation programme (grant agreement No.~948381).
\item Competing interests: The authors declare that there are no competing interests.
\item Ethics declaration: Not applicable.
\item Data availability: This work is based on publicly available data from the SLSN Catalog (\url{https://github.com/gmzsebastian/SLSNe}).
\end{itemize}








\bibliography{ptf12dam-oii}

\end{document}